\documentclass[conference]{IEEEtran}
\IEEEoverridecommandlockouts
\usepackage{cite}
\usepackage{amsmath,amssymb,amsfonts}
\usepackage{placeins}
\usepackage{algorithmic}
\usepackage{graphicx}
\usepackage{textcomp}
\usepackage{xcolor}
\usepackage{subcaption}
\usepackage{lineno,wrapfig,graphicx,amssymb,enumerate}
\usepackage{caption}
\usepackage{multirow,bigstrut}
\usepackage{booktabs}
\usepackage[linesnumbered,ruled,vlined]{algorithm2e}
\usepackage{float}
\begin{document}

% \title{Subject Agnostic Facial Inpainting Using DCGAN with Perceptual and Contextual Losses}

\title{Modelling Social Context for Fake News Detection: A Graph Neural Network Based Approach}

% Graph Neural Network based Propagation model for Fake News Detection

% Fake News Detection based on News Content and Context of Tweets using Graph Neural Network

%\author{\IEEEauthorblockN{Anonymous Authors}}

\author{\IEEEauthorblockN{Pallabi Saikia\textsuperscript{1}, Kshitij Gundale\textsuperscript{1}, Ankit Jain\textsuperscript{1}, Dev Jadeja\textsuperscript{1}, Harvi Patel\textsuperscript{1}, Mohendra Roy\textsuperscript{2}}
%\email{\{Pallabi.Saikia, Kshitij.gce18, Ankit.jce18, Dev.jce18, Harvi.pmtds20, Mohendra.Roy\}@sot.pdpu.ac.in}
\IEEEauthorblockA{\textit{\textsuperscript{1}Computer Science and Engineering Department, School of Technology} \mbox{}\\  \textit{\textsuperscript{2}Information and Communication Technology, School of Technology} \mbox{}\\   \textit{Pandit Deendayal Energy University, Gandhinagar-382007, India} \mbox{}\\  \textit{*Corresponding Authors: mohendra.roy@ieee.org; pallabi.iitg@gmail.com} }}

%*Corresponding Authors: mohendra.roy@ieee.org; pallabi.iitg@gmail.com

\maketitle

\begin{abstract}

Detection of fake news is crucial to ensure the authenticity of information and maintain the news ecosystem's reliability. Recently, there has been an increase in fake news content due to the recent proliferation of social media and fake content generation techniques such as Deep-Fake. The majority of the existing modalities of fake news detection focus on
content-based approaches. However, most of these techniques
fail to deal with ultra-realistic synthesized media produced
by generative models. Our recent studies find that the
propagation characteristics of authentic and fake news are
distinguishable, irrespective of their modalities. In this regard,
we have investigated the auxiliary information based on social
context to detect fake news. This paper has analyzed the social
context of fake news detection with a hybrid graph neural
network-based approach. This hybrid model is based on integrating a graph neural network on the propagation of news
and bi-directional encoder representations from the transformers
model on news content to learn the text features. Thus this
proposed approach learns the content as well as the context
features and hence able to outperform the baseline models
with an f1-score of 0.91 on Politifact and 0.93 on the Gossipcop
dataset, respectively

\footnote{\copyright IEEE, Paper No: 834, IJCNN, 2022 IEEE World Congress on Computational Intelligence }
% Firstly, we have demonstrated how the social context of fake news differ from real news in terms of size, number of followers and followings. Then we design a hybrid deep learning approach by building a graph neural network on propagation graph that learns on the context features, and BERT model on news content to learn on the text features. The proposed hybrid approach learns the content and the context information effectively and outperformed over the baseline models with an accuracy of 90\% on Politifact and 92\% on Gossipcop, respectively.

% The social context during news dissemination process on social media forms the inherent tri-relationship, the relationship among publishers, news pieces, and users, which has potential to improve fake news detection.

% show that real news are significantly bigger in size, are spread by users with more followers and less followings, and are actively spread on Twitter for a longer period of time than fake news. Secondly, we achieve an 87\% accuracy
% using a Random Forest Classifier solely trained on propagation features.
% Lastly, we design a Geometric Deep Learning approach to the problem by
% building a graph neural network that directly learns on the propagation
% graphs and achieve an accuracy of 73.3\%.

\end{abstract}

\section{Introduction}

% Social media has revolutionized the way people access information. 

Social platforms like Facebook and Twitter are becoming increasingly popular for day-to-day news consumption due to ease of access, low cost, and fast news dissemination \cite{PRC}. As a result, these platforms have increasingly become the dominant source of information. However, the authenticity of news on these platforms is suspicious without any regulatory mechanism. Hence, social media also enable the wide propagation of fake news, implanted with false information. False or misleading information, commonly known as fake news, can cause significant damage to individuals and society. A series of recent incidents have demonstrated the potential of fake news in damaging personal, economic and national integrity. For example, a recent rumor caused the death of 800 people after the consummation of alcohol-based cleaning products as a cure for Covid-19 \cite{world-53755067}, and the influence of fake news in the democratic process of a country \cite{marwick2017media}. Looking at these consequences of fake news, many organizations have considered it a global challenge.  

Fake news can be classified as parody, satire, fabricated news, propaganda, etc. \cite{tandoc2018defining}. Moreover, the term also confine the concepts of disinformation (intentionally misleading information), misinformation (information that can be proven to be false), manipulation, and rumors \cite{lazer2018science}. Although many definitions of fake news exist, there is no universally accepted one. One generalized definition of fake news introduced in literature \cite{zhou2020survey} is "fake news is intentionally and verifiably false news published by a news outlet."

% Therefore, how to automatically and accurately identify fake news before it is widespread has become an urgent challenge for research. Many groups are taking actions against the fake news diffusion, but a systematic way to detect them on social media is still an ongoing research. 
% Content-based approaches make use of news article's headline and body content to determine whether news is fake or real, whereas propagation-based approaches take into consideration the propagation pattern of news. This includes the interactions between users and tweets by the means of follow, retweet and like functionalities on twitter. 

Traditional sources of media such as television and newspapers have a structure of one-to-many. However, with its millions of monthly active users, social platforms such as Twitter are examples of a many-to-many approach. Therefore, surveillance of information diffused in such platforms is relatively complicated. Moreover, news on social platforms is emerging unprecedented, making it increasingly difficult to fact-check. Many fact-checking websites such as Snopes \cite{snopes}, and Politifact \cite{politifact} exist to combat the problem of fake news. However, most of these are purely on manual methods and are difficult to scale up. Therefore, researchers are now focusing on data-driven or machine learning-based approaches to automatically and accurately detect fake news. Most of these approaches are based on user and content-based features, which are insufficient to address state-of-art generative modalities. However, a recent study shows that the propagation characteristics of news vary based on their nature, irrespective of their content properties \cite{vosoughi2018spread}. Therefore, features corresponding to the propagation patterns of news could be effectively used as a basis for fake news detection on social media  \cite{han2020graph} \cite{meyers2020fake} \cite{silva2021propagation2vec}. Propagation patterns are useful in incorporating the context of social influence, but in contrast to content-based features, the features have the advantage of being language and content-agnostic. 

% Content-based features include headline of news article, body content, whereas propagation-based features include the features like the interactions between users and tweets, number of followers, retweet, etc. Only a few studies so far have leveraged these features for the fake news detection task \cite{han2020graph} \cite{meyers2020fake} \cite{silva2021propagation2vec}. Additionally, most of them have based their analysis on a narrow tweet retrieval methodology that only considers tweets to be propagating a news piece if they explicitly contain an URL link to an online news article. 

This paper explores a method of constructing the propagation graph of the social media network, following the propagation structure of Twitter posts. We then explored the graph neural network-based representation learning algorithm to extract the propagation features from the structured graph automatically. A hybrid model is built that exploits both the context features extracted with the graph neural network and the content features with the transformer model and embeds both the textual and structural information into a high-level abstract representation that can be effective for better analysis of the propagated tweet in a social network. We empirically evaluate our proposed model on two public datasets from Twitter, Politifact, and Gossipcop. The proposed model is compared with the baseline models on multiple evaluation indicators such as Accuracy, Precision, Recall, F1-score, and AUC. Moreover, we have also analyzed and compared the performance of the proposed model with the model based on manually extracted propagation characteristics of news for characterization of its authenticity. The details of the proposed approach are discussed in the following sections.

\section{Background and Related Work}

% This section reviews prior work in the field of fake news detection.

Fake news detection has gotten much attention in recent years as a research subject. The existing approaches of fake news detection in the literature typically are of three types (Shu et al. (2017)) \cite{shu2017fake}, namely news-based, user-based and propagation-based, which are based on the use of different types of information available in the social media. Moreover, news-based approaches fall into the category of content-based approaches, whereas the other two approaches lead to context-based. Content-based and Context-based approaches merged are also getting popular nowadays for effective detection of propagation of fake news, which is also called the mixed approach   \cite{silva2021propagation2vec}.

Content-based approaches attempt to solve the problem of fake news classification by using the news article's headline and body; hence it is called content-based. The underlying idea of this kind of approach is that fake news exhibit a significantly different presentation style than real news. In a work presented by Perez et al. \cite{perez2018automatic}, different text-based content features, namely, Ngrams, Punctuation, Psycholinguistics features, readability, and syntax, are extracted from the text of the news. The linear SVM model was trained on such features. In another work presented by Horne et al. \cite{horne2017just}, similar feature engineering was applied but considered satire as one of the classes along with fake and real. In work proposed by Nor et al. \cite{norregaard2019nela}, weak labeling was utilized, where labeling of news articles was done based on which category their source belonged to. The considered features extracted from the news content are Style, Complexity, Bias, Affect, Moral, and Event. Wang et al. \cite{wang2017liar} experimented on the Politifact dataset and used six unique labels for the target variable. Traditional ML algorithms were trained mainly focusing on neural nets, which provided promising results over all the considered models.

Context-based approaches mainly rely on propagation patterns of news on social media networks (e.g., Twitter) to classify news articles. Propagation patterns are constructed by considering interactions between tweets and users' following, retweets, and likes. In work proposed by Wu et al. \cite{wu2015false}, a hybrid kernel function based on a random walk graph kernel and an RBF kernel using propagation features was proposed to model the propagation behavior of fake news. Fake news spreads can be easily modeled as graphs on social media platforms, and Graph Neural Networks recently have been popular for automatically extracting propagation features of graphs and designing better models for fake news detection \cite{monti2017geometric, wu2020comprehensive, meyers2020fake, han2020graph}. Propagation patterns have the distinct advantage of being language and content-agnostic. Comparatively limited studies \cite{han2020graph} \cite{meyers2020fake} \cite{silva2021propagation2vec} have been found in leveraging propagation features for detection of fake news. The graph classification approach aims to optimize the use of propagation features, and the success of graph neural network approaches in the prior works \cite{han2020graph, zhou2018fake, zhou2018fake, liu2018early} motivates further investigation of the graph neural network model for characterization of fake news.

% However, state-of-the art results of this approach over many other domains of structured data, including fake news, motivates the researchers of this domain for further investigation of the graph neural network model for characterisation of fake news. demonstrates the value of using propagation features to distinguish between real and fraudulent news.

% Moreover, most of the approaches are based on a narrow tweet retrieval analysis that considers social post to be propagating a news piece if they explicitly contain an URL link to an online news article. 

Mixed approaches are getting the most attention nowadays to combat the limitations of both content and context-based approaches by combining their advantages. The mixed approach uses both propagation pattern and content, usually in the text, to verify the validity of news articles. Ruchansky et al. \cite{ruchansky2017csi} proposed a model consisting of multiple components to extract representations of articles and to extract representations of users. Both these components are then integrated to get a resultant vector to be used for the final classification task. Kai et al. \cite{shu2019beyond} proposed a framework consisting of five components: a semi-supervised classification component, a publisher-news relation embedding component, a user embedding component, a news content embedding component, and a user-news interaction embedding component. The latent representations for news content and users are learned via non-negative matrix factorization, and the problem is then formalised as an optimization over the above components. In work proposed by Nguyen et al. \cite{nguyen2020fang}, a propagation graph was built using news sources, news articles, social users, and interactions between two entities at a given time.
Additionally, the stance of the tweet with respect to the news title was also taken into account. Bi-directional LSTM (Bi-LTSM) was used to optimize the fake news detection objective. The approach emphasizes learning generalization representations for social entities by optimizing three concurrent losses, namely, Stance loss, Proximity loss, and Fake News Detection Loss.

\section{Dataset description and Pre-processing} \label{dataset-desc}

\begin{figure}[!b]
\centering
\begin{subfigure}[b]{0.24\textwidth}
\centering
\includegraphics[width=\textwidth]{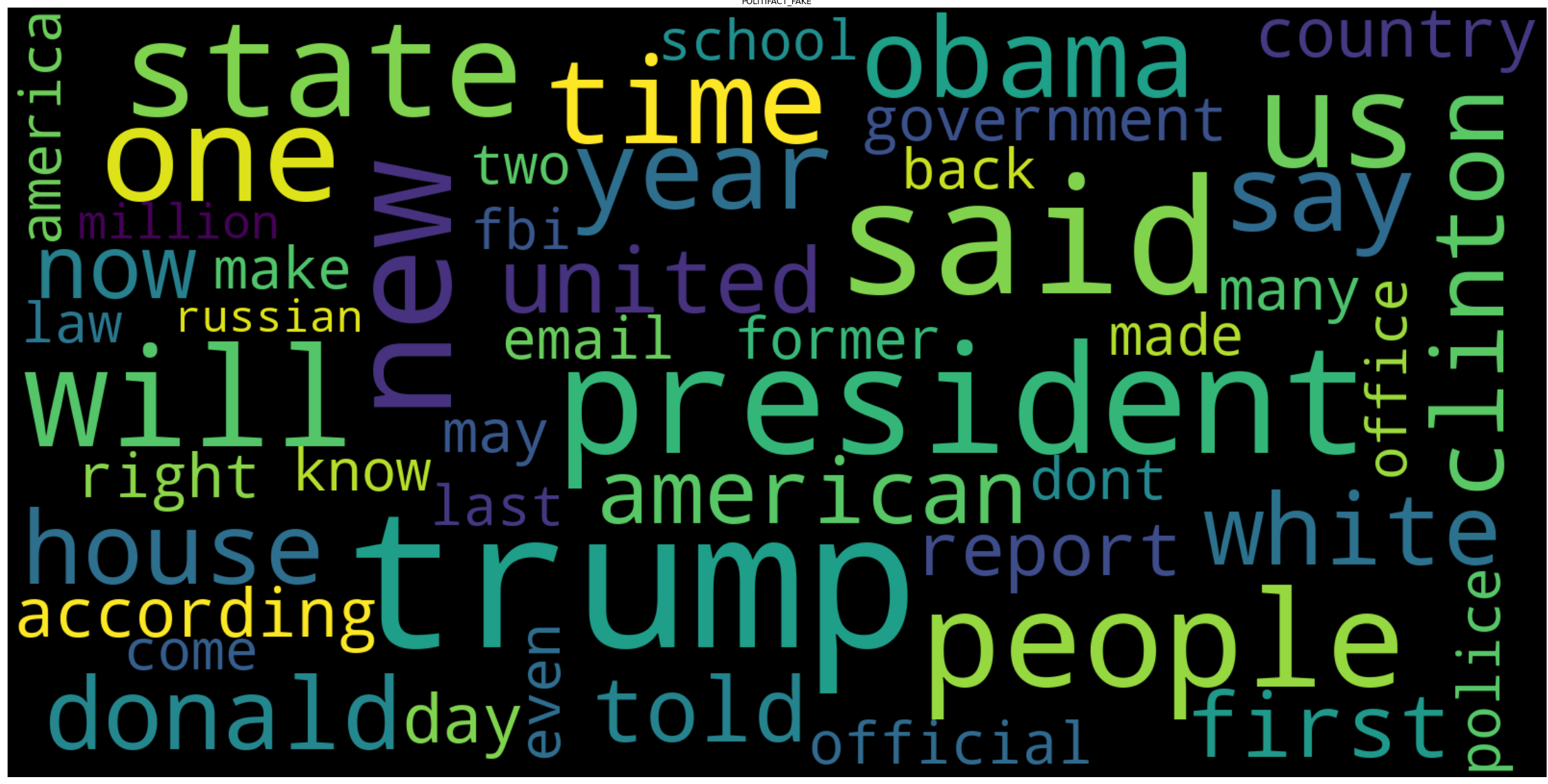}
\caption{Politifact Fake}
\end{subfigure}
\begin{subfigure}[b]{0.24\textwidth}
\centering
\includegraphics[width=\textwidth]{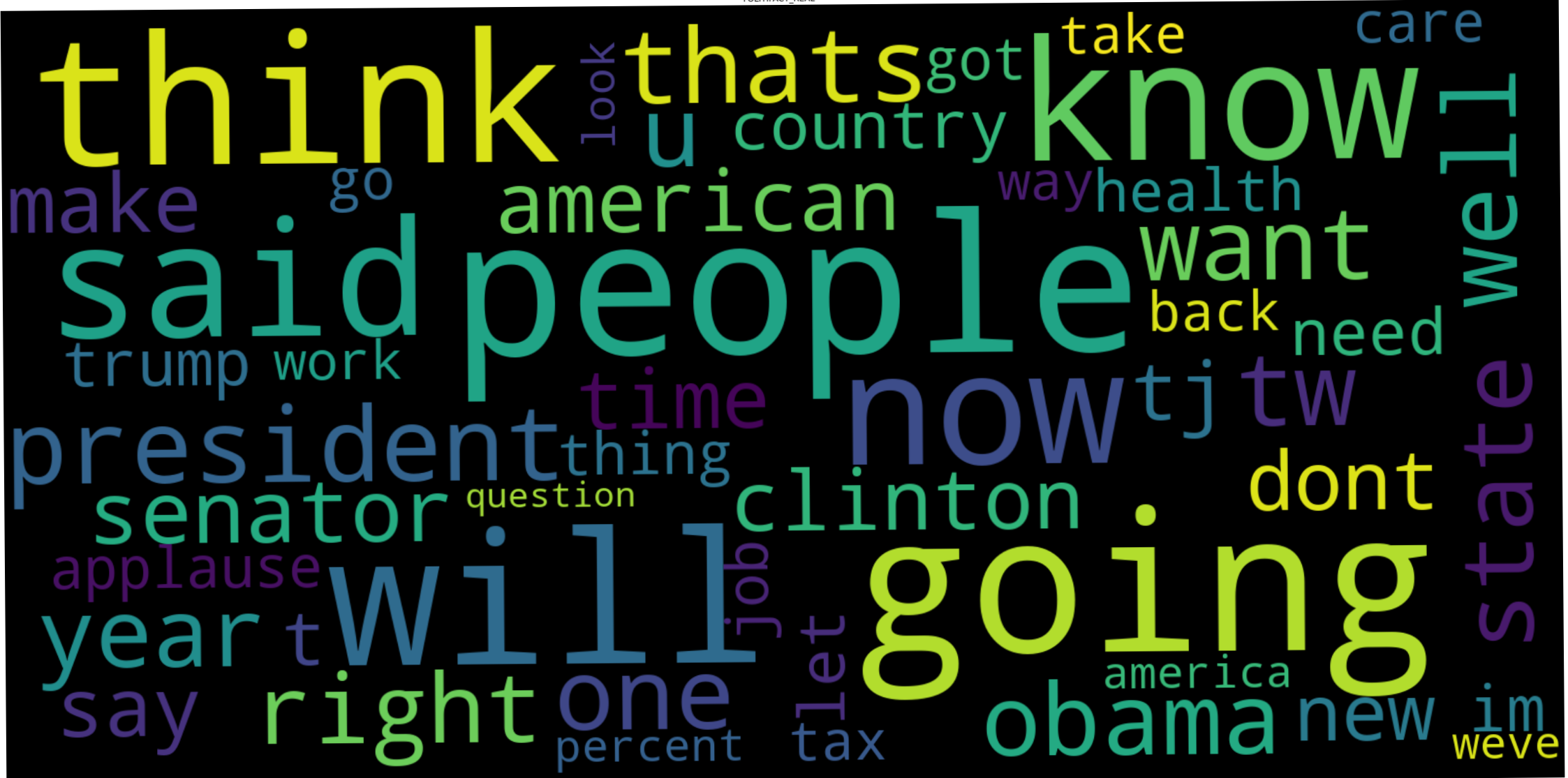}
\caption{Politifact Real}
\end{subfigure}
\begin{subfigure}[b]{0.24\textwidth}
\centering
\includegraphics[width=\textwidth]{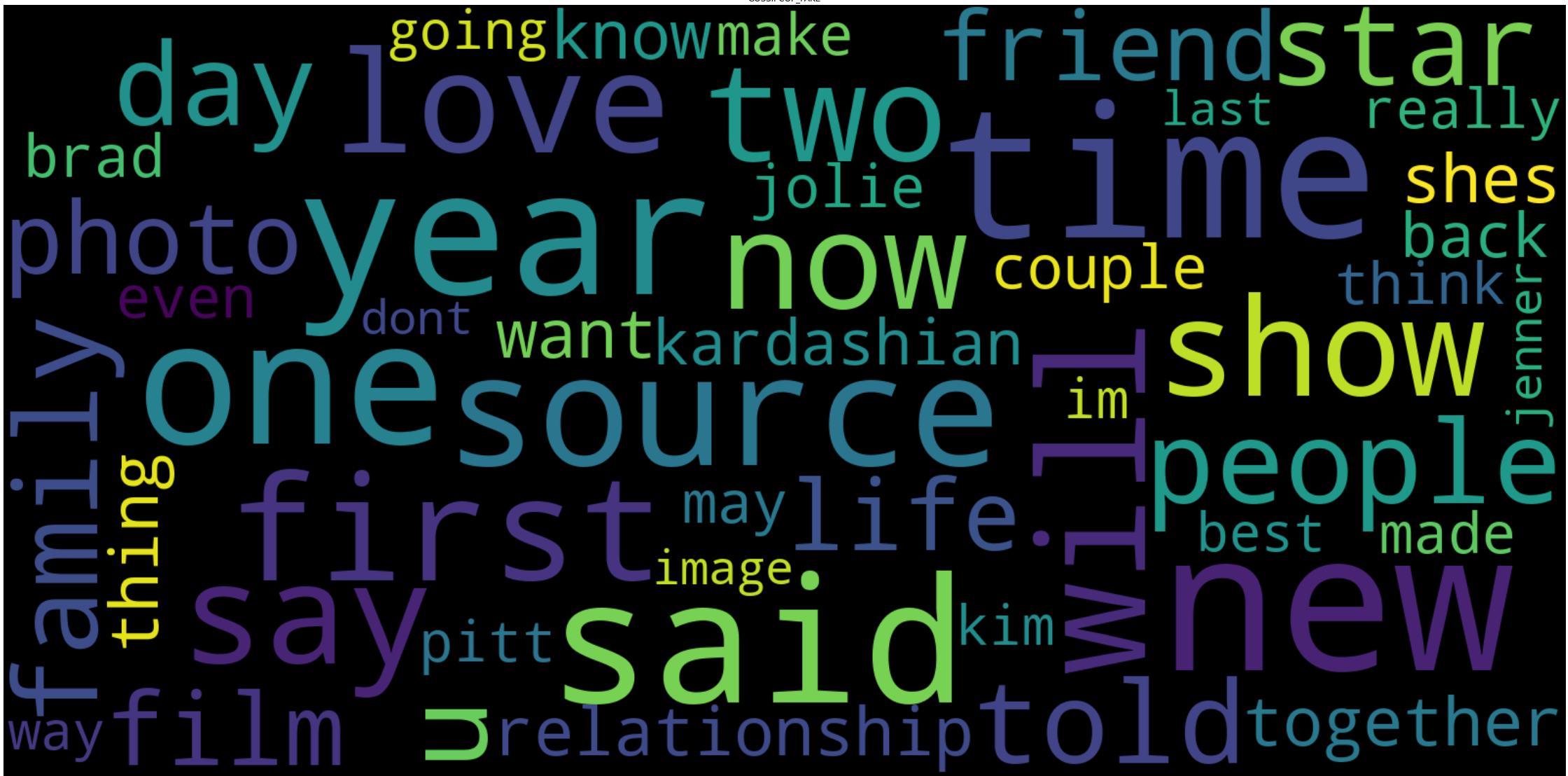}
\caption{GossipCop Fake}
\end{subfigure}
\begin{subfigure}[b]{0.24\textwidth}
\centering
\includegraphics[width=\textwidth]{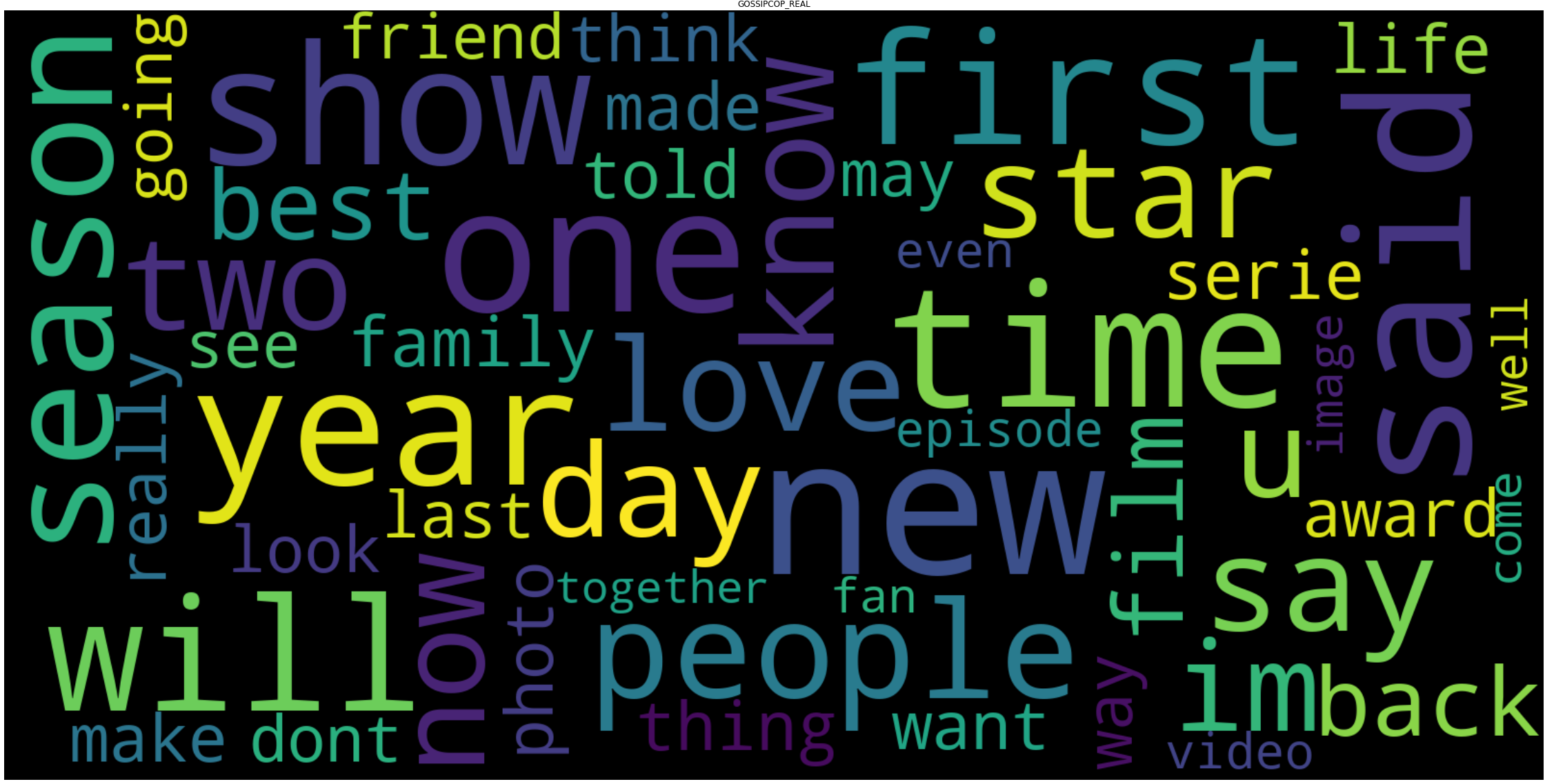}
\caption{GossipCop Real}
\end{subfigure}
\caption{Word clouds of news content}
\label{fig:word_clouds}
\end{figure}

% \subsection{Data Description}

In this work, we have used a public data repository, FakeNewsNet \cite{shu2018fakenewsnet}. The repository consists of comprehensive datasets from two popular fact-checking websites, Politifact and GossipCop. The datasets include social context, news content, and other dynamic information for the fact-checking websites. Politifact \cite{politifact} project is based on U.S. politics that reports on accuracy of statements made by elected officials, their staffs, lobbyists, candidates, interest groups and many others involved in. Whereas, GossipCop \cite{gossipcop} website fact-checks celebrity reporting. Both datasets have a number of articles whose ground truth is provided by its source (i.e. assigned by independent journalists). The word clouds representations of both the types of news are provided in Figure \ref{fig:word_clouds}.

\subsection{Data Collection and Pre-preprocessing}

% \cite{shu2018fakenewsnet} provides list of relevant tweet ids for every news article in the dataset. We collected following information using the script provided by authors of \cite{shu2018fakenewsnet} \textbf{Cite github repo? Zenedo?} - news articles, tweets, retweets and user profiles. 

We extracted the available information of the news based on the approach mentioned in work \cite{shu2018fakenewsnet}: news body, tweets, retweets and user profiles, that is relevant to tweet ids for every news article in the datasets. For every news article in the dataset relevant tweet ids are available. However, we discarded the news articles with missing text content. Some statistics of the collected datasets are provided in Table \ref{tab:Data}. 

\begin{center}
\begin{table}[!h]
\caption{Dataset Statistics}
\label{tab:Data}
\resizebox{0.47\textwidth}{!}{%
\begin{tabular}{|c c c c c|}
\hline
\multirow{2}{*}{}    & \multicolumn{2}{l}{Politifact}        & \multicolumn{2}{l|}{Gossipcop}         \\ \cline{2-5} 
                     & \multicolumn{1}{l}{Fake}    & Real    & \multicolumn{1}{l}{Fake}    & Real    \\ \hline
News Articles        & \multicolumn{1}{l}{432}     & 624     & \multicolumn{1}{l}{5323}    & 16817   \\ \hline
Tweets               & \multicolumn{1}{l}{164,892} & 399,237 & \multicolumn{1}{l}{519,581} & 876,967 \\ \hline
Unique Users         & \multicolumn{1}{l}{201,748} & 596,435 & \multicolumn{1}{l}{504,638} & 199,031 \\ \hline
\end{tabular}%
}
\end{table}
\end{center}

\subsection{Propagation Graph Construction}

For every news article in the dataset, Twitter API was used to retrieve its Tweets and Retweets. Propagation graph is then constructed to model how information disseminates from one user to another. However, Twitter API does not provide an immediate source of a retweet. For example, \(tweet_0\) has been retweeted in \(retweet_1\). If \(retweet_1\) is retweeted again in \(retweet_2\), twitter would store \(tweet_0\) as source of both retweets. In order to determine the immediate source of a retweet, all tweets and retweets in the given set are sorted based on timestamp. Let \{\(tweet_0\),  \(retweet_1\),  \(retweet_2\) ...\} be a set of sorted tweets and retweets. For each \(retweet_i\) , its immediate source is searched from tweets/retweets in the same set which were published earlier using the following heuristics, 
\begin{enumerate}
\item \(retweet_j\) is identified as source of \(retweet_i\) if owner of \(retweet_j\) mentions owner of \(retweet_i\).
\item Or if \(retweet_i\) is published within a certain period of time after \(retweet_j\). 
\item Else  \(tweet_0\) is considered as source of \(retweet_i\) .
\end{enumerate}

% From those labelled news articles, the headline is extracted and separated into a set of keywords. Then, those keywords are concatenated into a query for the Twitter API. For each news article, labelled real or fake, different kinds of information are then accessed:

% \begin{enumerate}
%     \item news content: the body of the article, images, publish date
%     \item tweets: the list of tweets containing the article headline keywords
%     \item retweets: the list of retweets of all tweets previously retrieved
%     \item user information: the profile information (user id, creation date, 200 most recent published tweets, list of followers and friends) of all users that have posted a tweet or retweet related to the news article.
% \end{enumerate}

% This data set provide us with the necessary information to create the propagation graphs detailed in the following section. 

\begin{figure}[!h]
\centering
\begin{subfigure}[b]{0.24\textwidth}
\centering
\includegraphics[width=\textwidth]{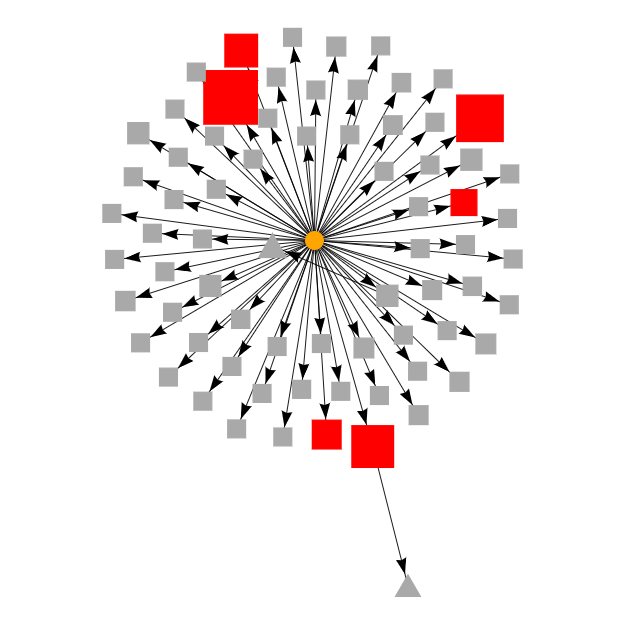}
\end{subfigure}
\begin{subfigure}[b]{0.24\textwidth}
\centering
\includegraphics[width=\textwidth]{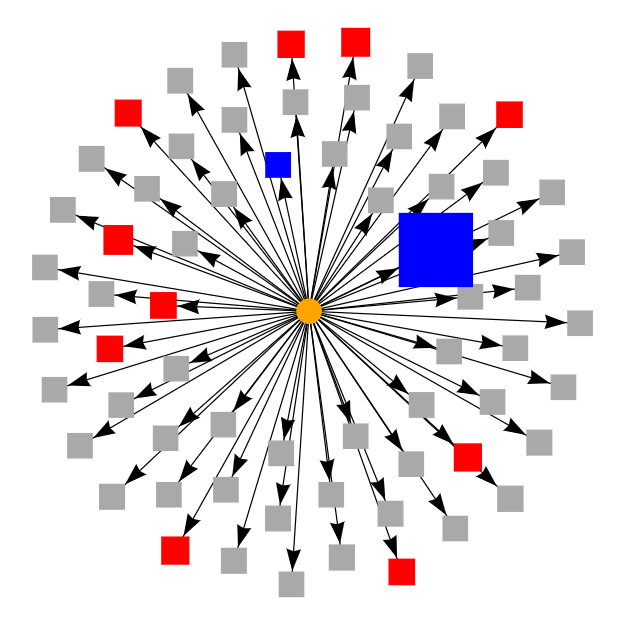}
\end{subfigure}
\begin{subfigure}[b]{0.24\textwidth}
\centering
\includegraphics[width=\textwidth]{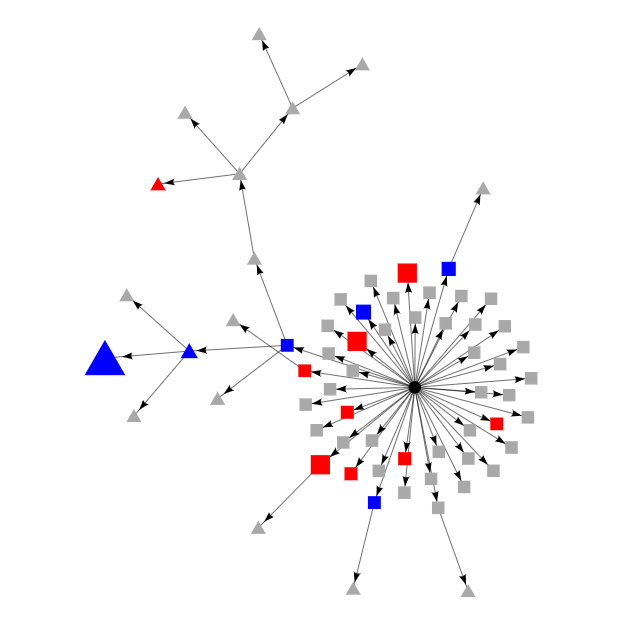}
\end{subfigure}
\begin{subfigure}[b]{0.24\textwidth}
\centering
\includegraphics[width=\textwidth]{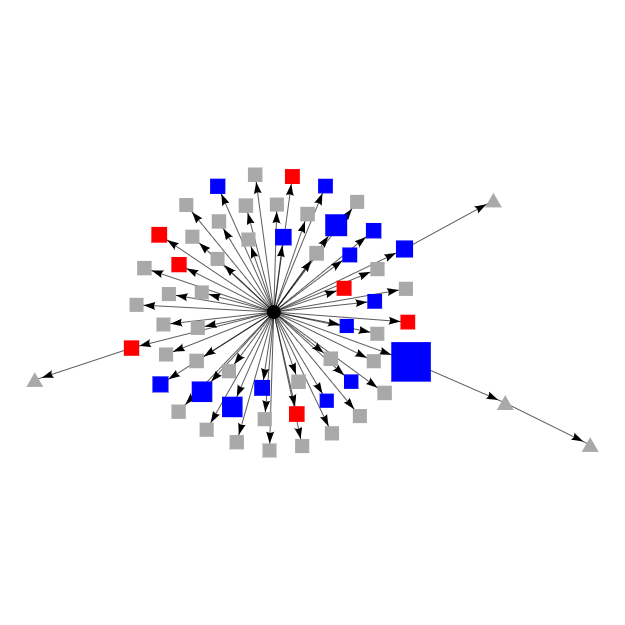}
\end{subfigure}
\caption{Yellow nodes indicate real news while Black nodes indicate fake news. Square are tweets and triangles are retweets of the particular news considered. Orange colored nodes indicate author of that node(tweet/retweet) has more than 10000 followers. Blue nodes indicate author of the node is verified. Size of a node is proportional to the number of followers of its author}
\label{fig:propagation_graph}
\end{figure}

Following the above mentioned heuristics return an information cascade for tweet. All such cascades are connected to the source news node to construct a propagation graph of the news articles as shown in Figure \ref{fig:propagation_graph}. Followers/Following information is not considered in construction of propagation graphs due to the strict Twitter API rate limits which would limit their availability at inference time.

% The propagation graphs, derived from the set of tweets and retweets corresponding to a labelled piece of information, are defined as follows:
% – Let V be the set of nodes of the graph. A node can be of two types:
% 1. A tweet node: the node stores the tweet and its associated user. A tweet
% belongs to a news graph if it contains the keywords extracted from the
% headline of the news article.
% 2. A retweet node: the node stores the retweet and its associated user. All
% retweets of a tweet node are present in the graph.
% – Let E be the set of edges of the graph. Edges are drawn between a tweet
% and its retweets. Edges contain a time weight that corresponds to the time
% difference between the tweet and retweet publish times.
% Then G = (V,E) is the news graph. G is then a composition of non-connected
% sub-graphs where each sub-graph comprises a tweet and its associated retweets.
% It is important to note that Twitter is designed in such a way that a retweet of
% a retweet will point back to the original tweet. Hence, the depth of the graph is
% never more than 1.

\subsection{Feature Engineering}

Feature extraction performed in this work is divided into two parts - node-level and graph-level. 

% Each node in a propagation graph is assigned with a set of node-level features which extracted on basis of its characteristics and neighborhood. Conversely, we also use statistics and aggregation to obtain a single representation for entire propagation graph.

\subsubsection{Node-level features}

% We follow the feature engineering process described in \cite{silva2021propagation2vec} 

%  We extracted three types of node level features - User-based, text-based and temporal as mentioned in Table \ref{Table 3}.

Each node in the propagation graph is either a tweet or retweet with a corresponding user. Node-level features are usually extracted on the basis of its characteristics and neighborhood. Different node level features are considered including user-based, text-based and temporal as mentioned in Table \ref{Table 3}. User-based features are basically the characteristics of author of node, which includes verified status, number of followers and number of friends. Friends in twitter terminology are accounts that a user follows. Text-based features are extracted from text content of the node (tweet/retweet) and try to capture the sentiment of the tweet. These include number of hastags, number of users mentioned, sentiment score using VADER\cite{hutto2014vader} and frequency of positive and negative words. The temporal features we collected include difference in publication time with source node, parent and neighbors, which take into account the timeline of the node and its neighbors. 

% Each node in the propagation graph is either a tweet or retweet with a corresponding user, three types of node level features are added as shown in \cite{silva2021propagation2vec} - User-based, text-based and temporal selected in accordance with the dataset available.
\FloatBarrier
\begin{table}[]
\centering
\caption{Node-level features}
\label{Table 3}
\resizebox{0.48\textwidth}{!}{%
\begin{tabular}{|l|l|}
\hline
Feature Class               & Feature Name                                          \\ \hline
\multirow{3}{*}{User Based} & Is verified user                                      \\ \cline{2-2} 
                            & Number of Friends                                     \\ \cline{2-2} 
                            & Number of Followers                                   \\ \hline
\multirow{4}{*}{Text Based} & Number of Hashtags                                    \\ \cline{2-2} 
                            & Number of mentions                                    \\ \cline{2-2} 
                            & Sentiment score computed using VADER                  \\ \cline{2-2} 
                            & Frequency of positive words                           \\ \hline
\multirow{5}{*}{Temporal}   & Frequency of negative words                           \\ \cline{2-2} 
                            & User account timestamp                                \\ \cline{2-2} 
                            & Time difference with source node                      \\ \cline{2-2} 
                            & Time difference with immediate predecessor            \\ \cline{2-2} 
                            & Average time difference with the immediate successors \\ \hline
\end{tabular}%
}
\end{table}

% \begin{table}[h!]
% \begin{tabularx}{0.4\textwidth}{|c c|l}
% \cline{1-2}
% \multirow{3}{*}{User-based} & Is verified user                                      &  \\ \cline{2-2}
%                             & Number of Friends                                     &  \\ \cline{2-2}
%                             & Number of Followers                                   &  \\ \cline{1-2}
% \multirow{5}{*}{Text-based} & Number of Hashtags                                    &  \\ \cline{2-2}
%                             & Number of mentions                                    &  \\ \cline{2-2}
%                             & Sentiment score computed using VADER                  &  \\ \cline{2-2}
%                             & Frequency of positive words                           &  \\ \cline{2-2}
%                             & Frequency of negative words                           &  \\ \cline{1-2}
% \multirow{4}{*}{Temporal}   & User account timestamp                                &  \\ \cline{2-2}
%                             & Time difference with source node                      &  \\ \cline{2-2}
%                             & Time difference with immediate predecessor            &  \\ \cline{2-2}
%                             & Average time difference with the immediate successors &  \\ \cline{1-2}
% \end{tabularx}
% \caption{Node-level features}
% \label{Table 3}
% \end{table}

\subsubsection{Graph-level features}

% \cite{silva2021propagation2vec} further builds upon this idea by proposing a iterative algorithm to update the aggregated features of each node and representing the graph using the aggregated features of source node.

A simple approach to extract a vector representation of a propagation graph would be applying aggregation techniques (averaging, max, min) on handcrafted node-level features \cite{meyers2020fake}. In this paper, we have used simple averaging of node-level features along with meta information of graph as proposed by Meyers et. al. \cite{meyers2020fake} as a baseline. We start with collection of basic statistics at graph-level - number of nodes, number of tweets and number of users. Next, mean aggregation is used to incorporate node-level information such as average number of friends, followers, retweets per tweets and time between tweet and its retweets. Finally, we collect temporal features such as total amount of time news article of referenced on twitter which is essentially the difference between the publication time of first and the last tweet, number of users involved in propagation after 10 hours of news publication and percentage of tweets/retweets published in first 60 minutes. Table \ref{tab:graph-level-features} shows different graph-level features collected.

\FloatBarrier
\begin{table}[h!]
\huge
\caption{Graph-level Features}
\label{tab:graph-level-features}
\resizebox{0.48\textwidth}{!}{%
\begin{tabular}{|l l|}
\hline
num\_nodes               & Total number of tweets and retweets for a news article       \\ \hline
num\_tweets              & Number of tweets                                             \\ \hline
avg\_num\_retweets       & Average number of retweets per tweet                         \\ \hline
retweet\_perc            & Average number of retweets per tweet                         \\ \hline
num\_users               & Number of unique users                                       \\ \hline
total\_propagation\_time & Amount of time news was referenced on twitter                \\ \hline
avg\_num\_followers      & Number of followers averaged over all users                    \\ \hline
avg\_num\_friends        & Number of friends averaged over all users                    \\ \hline
perc\_posts\_1\_hour     & Percentage of tweets in first hour of news publish time \\ \hline
users\_10h               & Users reached in 10 hours                                    \\ \hline
avg\_time\_diff          & Avg. time between a tweet and its retweet      \\ \hline
\end{tabular}%
}
\end{table}

% check architecture here https://dl.acm.org/doi/pdf/10.1145/3366423.3380083?casa_token=AdgM0bH2cl0AAAAA:1-bL8PJ4m2Yz

% \begin{figure*}[!h]
% \centering
% \includegraphics[width=0.95\textwidth]{images/early-fusion-arch.png}
% \caption{Early Fusion architecture}
% \label{fig:early-fusion-arch}
% \end{figure*}

% \begin{figure*}[!h]
% \centering
% \includegraphics[width=0.95\textwidth]{images/late-fusion-arch.png}
% \caption{Late Fusion architecture}
% \label{fig:late-fusion-arch}
% \end{figure*}

%\FloatBarrier
%\begin{figure*}[h!]
 % \centering
 % \begin{subfigure}{\linewidth}
  %  \includegraphics[width=\linewidth]{early-fusion-arch}
   % \caption{Early Fusion architecture}
  %  \label{fig:early-fusion-arch}
  %\end{subfigure}%
  %\vfill
  %\begin{subfigure}{\linewidth}
   % \includegraphics[width=\linewidth]{late-fusion-arch}
%    \caption{Late Fusion architecture}
 %   \label{fig:late-fusion-arch}
 % \end{subfigure}%
  %\caption{Module to combine content and context features.}
 % \label{fig:something}
%\end{figure*}

\begin{figure*}[!htb]
\centering
\includegraphics[width=0.9\linewidth]{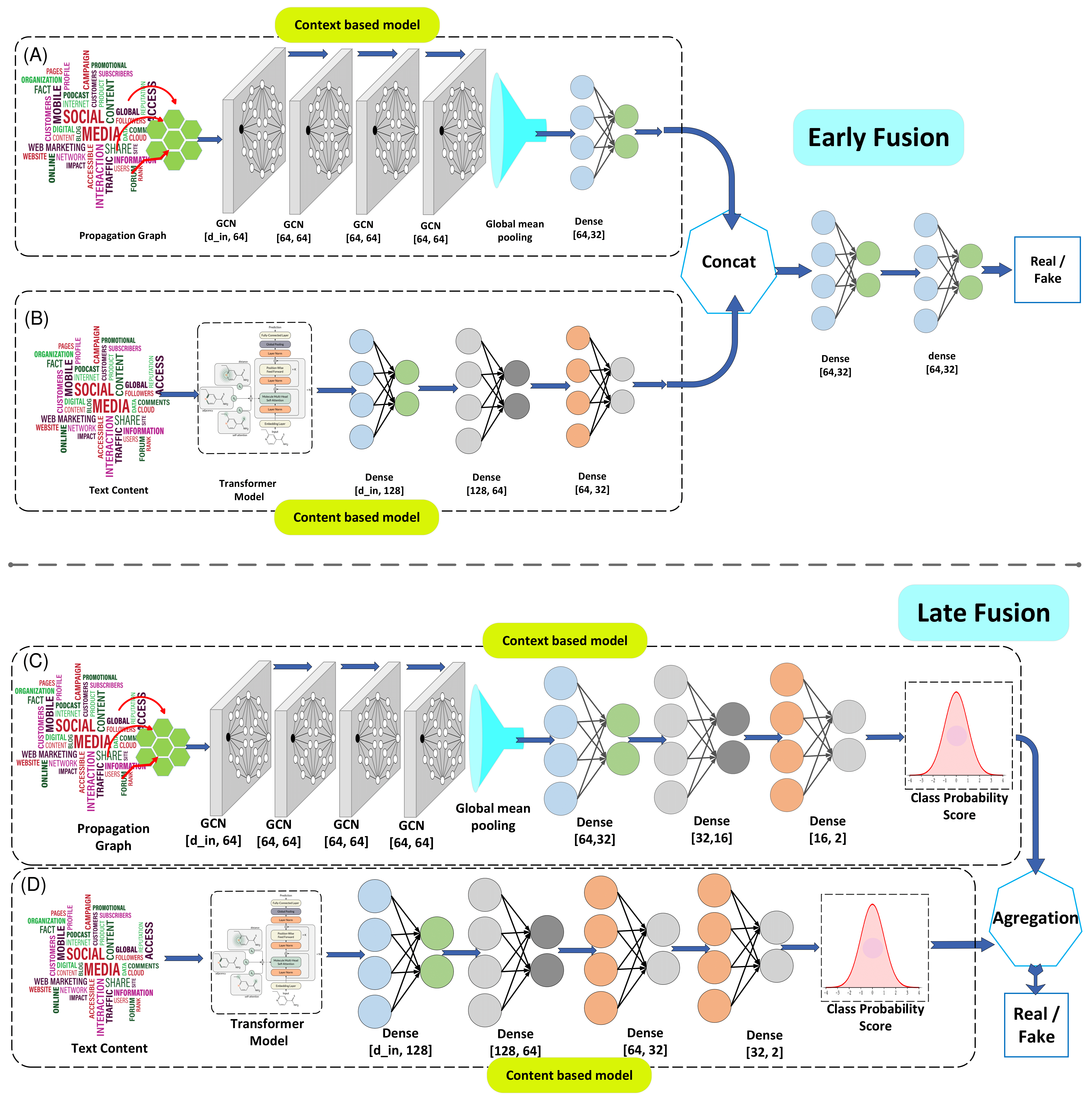}
\caption{Proposed Workflow of early and late fusion methods.}
\label{fig:ELT}
\centering
\end{figure*}

%---------------
\section{Methodology} \label{methodology}

Our proposed methodology incorporates basically three modules: (a) Graph Neural Network for modelling propagation context of news, (b) Pretrained Transformer model to learn from the news content, and (c) finally a mixed module to combine the representation of the above two modules.

\subsection{Graph Neural Network for modelling propagation pattern}

Graph Neural Networks (GNN) are a class of neural network that operates directly on graph structures. Social media platforms, like twitter, can be modelled as graphs, as shown in Figure \ref{fig:propagation_graph}. GNN work on the principle of message passing or neighborhood aggregation which is a iterative process to generate node embeddings by aggregating information from local neighborhood. Consider a graph \begin{math}G = (V,E)\end{math} with a corresponding node feature matrix \(X \in \mathbb{R}^{d\times|V|}\) and adjacency matrix \(A \in \{0,1\}^{|V|\times|V|}\). After \(k\) iterations of message-passing, the node embeddings can be represented by,

\begin{equation}
H^{(k)} = f(A,H^{(k-1)}; W^{(k)})
\end{equation}

where \(H^{(k)}\) is node embedding matrix or output of GNN after $k$ iterations, \(f\) is message passing function with trainable parameters \(W^{(k)}\). GNNs aggregate the neighbourhood representation within $k$ hops and then apply a pooling such as $SUM$, $MEAN$, $MAX$ to obtain the final representation of the node. The representation which incorporates the social context information can then be used to classify the graphs. The general steps involved in training of GNN involves:

\begin{enumerate}
    \item Generate node embeddings using multiple iterations of message passing
    \item Extract graph embedding by aggregation of node embedding
    \item Feed the embeddings into fully-connected layers for classification
\end{enumerate}

Our work is based on building on the apparent potential of abstract features extracted by GNN on propagation network of twitter to detect fake news. The working principle can be defined as: 
Given a news propagation graph $G$ of a specific news item, that consists of a sets of tweets and retweets, how significant the propagation features are at classifying the news $G$ as fake or real. We applied here few most recent networks of GNN, namely GCNConv\cite{kipf2016semi}, GATConv\cite{velivckovic2017graph} and GraphConv\cite{morris2019weisfeiler} for modelling propagation behaviours of fake news. The details of each of model are discussed below:

\vspace{2mm}
 
\noindent
\textbf{GCNConv}: GCNConv \cite{kipf2016semi} is a graph-based semi-supervised learning algorithm outlined in where the learner is provided with an adjacency matrix, $A$, and node features, $X$, as input and a subset of node labels, $Y$, for training. GCN is spectral based, where eigen-decomposition of the graph Laplacian is used in network propagation. This spectral method is used to aggregate neighboring nodes in a graph to infer the value of the current node. In GCNConv, eigen-decomposition is performed via approximation to reduce runtime. The propagation rule of GCNConv, can be represented by the following equation:

\begin{equation}
    \mathbf{X}^{\prime} = \mathbf{\hat{D}}^{-1/2} \mathbf{\hat{A}}
    \mathbf{\hat{D}}^{-1/2} \mathbf{X} \mathbf{\Theta},
\end{equation}

where, \begin{math}\mathbf{\hat{A}}=\mathbf{A}+\mathbf{I}\end{math} indicates the adjacency matrix with self-loops for every nodes, $X^{\prime}$ is intermediate node embedding matrix or output after applying message passing function on embedding $X$ with layer specific trainable parameters $\Theta$, $D$ is the degree matrix to normalise large degree nodes. \begin{math}\hat{D}{ii} = \sum_{j} \hat{A}_{i,j}\end{math} is the corresponding diagonal degree matrix that acts as a normaliser to circumvent numerical instabilities. The adjacency matrix $A$ consists of edge weights via the optional edge\_weight tensor. The node-wise formulation is provided below:

\begin{equation}
        \mathbf{x_i}^{\prime} = \mathbf{\Theta}^{\top} \sum_{j \in
        \mathcal{N}(v) \cup \{ i \}} \frac{e_{j,i}}{\sqrt{\hat{d}_j
        \hat{d}_i}} \mathbf{x}_j
\end{equation}

where, $N(v)$ is the neighboring nodes of node $i$, \begin{math}\hat{d_i} = 1 + \sum_{j \in \mathcal{N}(i)} e_{j,i}\end{math}, where \begin{math}e_{j,i}\end{math} denotes the edge weight from source node \begin{math}j\end{math} to target node \begin{math}i\end{math}.

% https://hkcornwell.github.io/docs/Project_Report_FEA_GCN.pdf
% https://hal.archives-ouvertes.fr/hal-02334445/document

\vspace{2mm}

\noindent
\textbf{GraphConv}: GraphConv \cite{morris2019weisfeiler} is a generalization of graph neural networks capable of taking into account higher order graph structures at multiple scales. The message passing function for GraphConv is given by,

\begin{equation}
        \mathbf{x}^{\prime}_i = \mathbf{\Theta}_1 \mathbf{x}_i +
        \mathbf{\Theta}_2 \sum_{j \in \mathcal{N}(i)} e_{j,i} \cdot
        \mathbf{x}_j
\end{equation}

% \begin{equation}
%         \mathbf{x}^{\prime}_i = \mathbf{\theta}_1 \mathbf{x}_i +
%         \mathbf{\Theta}_2 \sum_{j \in \mathcal{N}(i)} e_{j,i} \cdot
%         \mathbf{x}_j
% \end{equation}

where $e_{j,i}$ denotes the edge weight from source node $j$ to target node $i$. 

\vspace{2mm}
\noindent
\textbf{GATConv}: GATConv \cite{velivckovic2017graph} is a attention-based graph neural network algorithm. It is an extension of GCNConv where instead of assigning same weight to each neighboring node, different weights are assigned through attention coefficients. This is achieved without use of expensive matrix calculations or prior knowledge of graph structure as provided below:

\begin{equation}
        \mathbf{x_i}^{\prime} = \alpha_{i,i}\mathbf{\Theta}\mathbf{x}_{i} +
        \sum_{j \in \mathcal{N}(i)} \alpha_{i,j}\mathbf{\Theta}\mathbf{x}_{j}
\end{equation}

where the attention coefficients $\alpha_{i,j}$ are computed as

% \begin{equation}
%         \frac{
%         \exp\left(\mathrm{LeakyReLU}\left(\mathbf{a}^{\top}
%         [\mathbf{\Theta}\mathbf{x}_i \, \Vert \, \mathbf{\Theta}\mathbf{x}_j]
%         \right)\right)}
%         {\sum_{k \in \mathcal{N}(i) \cup \{ i \}}
%         \exp\left(\mathrm{LeakyReLU}\left(\mathbf{a}^{\top}
%         [\mathbf{\Theta}\mathbf{x}_i \, \Vert \, \mathbf{\Theta}\mathbf{x}_k]
%         \right)\right)}
% \end{equation}

\begin{equation}
        \alpha_{i,j}=\frac{
        \exp\left(\mathrm{LeakyReLU}\left(e_{ij}\right)\right)}
        {\sum_{k \in \mathcal{N}(i) \cup \{ i \}}
        \exp\left(\mathrm{LeakyReLU}\left(e_{ij}\right)\right)}
\end{equation}

% \begin{figure}[!h]
% \includegraphics[width=0.48\textwidth]{images/gnn-arch.png}
% \caption{Graph Neural Network architecture}
% \label{fig:gnn-arch}
% \end{figure}

Where, $e_{ij}$ denotes the importance of node $j$’s features to node $i$, and $N$ is the Neighbourhood of node $i$. 

% Due to the varied structure of graphs, nodes can have a different number of neighbours. To have a common scaling across all neighbourhoods, the Attention coefficients are Normalized.

\subsection{News content representation}

% We use the implementation provided by Gensim \cite{rehurek2011gensim} in our experiments.

Text content of a news article can provide important signals in distinguishing fake and real news. We adopt two approaches to get vector representation of text content, Doc2Vec\cite{le2014distributed} and Embeddings from pretrained transformer models \cite{reimers-2019-sentence-bert}. Doc2Vec is an unsupervised algorithm and extension of Word2Vec \cite{mikolov2013efficient} that computes vector representation of variable length documents. The difference between Word2Vec and Doc2Vec is the addition of a special token called Document ID which learns the vector representation of entire document. Another approach that we considered for encoding text content was making use of pretrained transformer models from \textit{sentence-transformers} library \cite{reimers-2019-sentence-bert}. Specifically, we considered (1) all-MiniLM-L12-v2 based on \cite{wang2020minilm} (2) all-distilroberta-v1 based on distilled version of \cite{liu2019roberta} (3) all-mpnet-base-v2 based on \cite{song2020mpnet}. Transformer networks output an embedding for each token in the input text which are then averaged to obtain fixed length embedding for the document. Since all the models that we considered had a maximum sequence length of 512 tokens (500 english words), we consider different parts of text as input when length of text is greater than 512 tokens. Specifically, we consider first 512 tokens, last 512 tokens and combination of first 256 and last 256 tokens. 

% \begin{figure}[!h]
% \includegraphics[width=0.48\textwidth]{images/text-arch.png}
% \caption{Text Model architecture}
% \label{fig:text-arch}
% \end{figure}

\subsection{Mixed Approach: Combine context and content features}

Research on multi-modal fusion has shown that models trained by combining data from multiple sources have a clear advantage over those trained using only one source \cite{sebastianelli2021paradigm, tardy2017fusion}. In our research we explore two fusion techniques to combine the content and context features - Early fusion and Late fusion. The mixed approach takes the benefits of both the modalities and hence can be assumed to be more effective. Early fusion based mixed approach is shown in Figure \ref{fig:ELT} involves concatenating vector representations of text content and propagation to be used as input to fully connected layers for classification. Dimensions of both modalities are reduced to 32 to prevent one modality from overwhelming the other modality. Conversely, Late fusion outputs final prediction by aggregating predictions from base-classifiers. We explore aggregation strategies for late fusion - mean and classifier-based. In Late Fusion mean approach, predictions from aggregated using simple while in classifier approach a meta-classifiers is trained on out-of-fold predictions of base-classifiers. Figure \ref{fig:ELT} also illustrates the late fusion architecture of our approach.

\section{Experimental Results and Analysis}

% In the following section, we evaluate the model based on the considered hypotheses and comparisons is made with other baseline approaches.

The experiments were conducted on Google Colab Pro with 25 GB RAM and the codes were developed using python 3. Different libraries are used for the experimentation are Pytorch-Geometric\cite{Fey/Lenssen/2019}, sklearn\cite{scikit-learn}, sentence-transformers\cite{reimers-2019-sentence-bert}, Gensim\cite{rehurek2011gensim}, and Pandas\cite{reback2020pandas}. The classification metrics used are Accuracy, F1-score, Precision, Recall.

% OpenCV, Keras, sklearn, Scipy, Pandas, face recognition. The classification metrics used are Accuracy, F1-score, Precision, Recall, AUC.
% We employ Adam optimizer with a learning rate of $1 \times 10^{-5}$. We used F1-score, Precision, Recall, AUC, Accuracy metric for evaluation. We have used VGG16, InceptionV3, ResNet50, Xception, MobileNetV2, and EfficientNetB7 baseline unimodals to perform experiments.

% \subsubsection{IDE}Google Colab
% \subsubsection{Runtime Environment}Hosted Runtime
% % \subsubsection{Runtime Type}GPU
% \subsubsection{RAM}12 GB
% \subsubsection{Programming Language}Python3
% \subsubsection{Datasets} Politifact, Gosipcop

% \begin{table}[!h]
% \caption{Different libraries used for experimentation}
% \begin{tabular}{ |p{2cm}|p{5cm}|  }
%  \hline
%  LIBRARIES & PURPOSE\\
%  \hline
% Pytorch-Geometric\cite{Fey/Lenssen/2019} &  Differents variants of messagepassing functions for GNNs.\\
% \hline
% % Keras & Used for model building\\
% % \hline
% sklearn\cite{scikit-learn} & Used for computing Metrics (Accuracy, Macro F1-score, Precision, Recall)\\
% \hline
% sentence-transformers\cite{reimers-2019-sentence-bert} & Pretrained transformer models \\
% \hline
% Gensim\cite{rehurek2011gensim} & Doc2Vec implementation \\
% \hline
% Pandas\cite{reback2020pandas}\cite{mckinney-proc-scipy-2010} & Used for file handling.\\
%  \hline
% \end{tabular}
% \label{tab:libraries}
% \end{table}

% Different libraries we employed for the experimentation are mentioned in Table \ref{tab:libraries}.

Dataset preparation (splitting and sampling) details for modelling are provided in Table \ref{tab:dataset-splits}. For Politifact, the train-test dataset ratio was kept at 4:1. For training of models, 90\% of samples are randomly selected for training from train dataset and remaining are used for validation. This process is repeated for 10 times and average of the considered evaluation metrics are reported. This strategy is used because of small size of Politifact dataset. For GossipCop, the dataset is split into train-test-val in ratio of 70:15:15. Cross validation is not used for GossipCop because of large size of dataset. In case of Late fusion classifier, 3-fold inner cross validation(CV) is used to generate out-of-fold predictions. Splitting of dataset is done in a stratified manner such that ratio of fake to real news remains same in all sets. Same splits are used for all experiments to ensure consistency and fair estimate of performance. Wherever applicable, each model is trained for maximum of 50 epochs and best model weights are selected from epoch with lowest validation loss. Learning is set to 0.001 and batch size of 64 is used.  

% We use Pytorch-Geometric\cite{Fey/Lenssen/2019} library to experiment on differents variants of messagepassing functions - GCNConv\cite{kipf2016semi}, GATConv\cite{velivckovic2017graph} and GraphConv\cite{morris2019weisfeiler}. We make use of functions from Scikit-Learn \cite{scikit-learn} to perform dataset splitting and preparation achieve this.
%\FloatBarrier
\begin{table}[h!]
\centering
\caption{Dataset Splitting}
\label{tab:dataset-splits}
\resizebox{0.45\textwidth}{!}{%
\begin{tabular}{|c|c|c|}
\hline
                 & \textbf{Politifact}  & \textbf{GossipCop} \\ \hline
Train-Test split & 80:20                & 85:15              \\ \hline
Train-Val split  & 10-fold CV           & 82.35:17.65        \\ \hline
Sampling         & Random over sampling & None               \\ \hline
Class Weights    & None                 & Uniform            \\ \hline
\end{tabular}%
}
\end{table}

%================================End of table 4===================

%================Beginning of Table 5=============================
%\FloatBarrier
\begin{table*}[h]
\centering
\caption{Classification scores using graph-level features}
\label{tab:graph-level-scores}
\resizebox{0.95\textwidth}{!}{%
\begin{tabular}{|c|rrrr|cccc|}
\hline
\multirow{2}{*}{Classifier} & \multicolumn{4}{c|}{Politifact}                                                                                                & \multicolumn{4}{c|}{GossipCop}                                                                            \\ \cline{2-9} 
                            & \multicolumn{1}{c|}{F1}       & \multicolumn{1}{c|}{Precision} & \multicolumn{1}{c|}{Recall}   & \multicolumn{1}{c|}{Accuracy} & \multicolumn{1}{c|}{F1}       & \multicolumn{1}{c|}{Precision} & \multicolumn{1}{c|}{Recall}   & Accuracy \\ \hline
PassiveAggressiveClassifier & \multicolumn{1}{r|}{0.394982} & \multicolumn{1}{r|}{0.475000}  & \multicolumn{1}{r|}{0.347086} & 0.601018                      & \multicolumn{1}{c|}{0.554933} & \multicolumn{1}{c|}{0.898990}  & \multicolumn{1}{c|}{0.496567} & 0.571099 \\ \hline
RidgeClassifier             & \multicolumn{1}{r|}{0.443599} & \multicolumn{1}{r|}{0.481250}  & \multicolumn{1}{r|}{0.289315} & 0.720028                      & \multicolumn{1}{c|}{0.678024} & \multicolumn{1}{c|}{0.874576}  & \multicolumn{1}{c|}{0.775107} & 0.742265 \\ \hline
LogisticRegression          & \multicolumn{1}{r|}{0.464448} & \multicolumn{1}{r|}{0.480000}  & \multicolumn{1}{r|}{0.325717} & 0.766790                      & \multicolumn{1}{c|}{0.680060} & \multicolumn{1}{c|}{0.882470}  & \multicolumn{1}{c|}{0.760515} & 0.738644 \\ \hline
SGDClassifier               & \multicolumn{1}{r|}{0.441990} & \multicolumn{1}{r|}{0.469565}  & \multicolumn{1}{r|}{0.330019} & 0.709389                      & \multicolumn{1}{c|}{0.697815} & \multicolumn{1}{c|}{0.894500}  & \multicolumn{1}{c|}{0.767811} & 0.752469 \\ \hline
ExtraTreesClassifier        & \multicolumn{1}{r|}{0.463866} & \multicolumn{1}{r|}{0.473913}  & \multicolumn{1}{r|}{0.341212} & 0.773682                      & \multicolumn{1}{c|}{0.875082} & \multicolumn{1}{c|}{0.933305}  & \multicolumn{1}{c|}{0.954936} & 0.913101 \\ \hline
RandomForestClassifier      & \multicolumn{1}{r|}{\textbf{0.484482}} & \multicolumn{1}{r|}{0.475000}  & \multicolumn{1}{r|}{0.403515} & 0.816466                      & \multicolumn{1}{c|}{\textbf{0.882283}} & \multicolumn{1}{c|}{0.935565}  & \multicolumn{1}{c|}{0.959657} & 0.918367 \\ \hline
\end{tabular}%
}
\end{table*}
%\FloatBarrier
%================================End of table 5===================

%================Beginning of Table 6=============================
%\FloatBarrier
\begin{table*}[h]
\centering
\caption{Classification scores using Graph Neural Networks}
\label{tab:gnn-results}
\resizebox{0.95\textwidth}{!}{%
\begin{tabular}{|c|cccc|cccc|}
\hline
\multirow{2}{*}{\begin{tabular}[c]{@{}c@{}}Convolutional \\ Layer\end{tabular}} & \multicolumn{4}{c|}{Politifact}                                                                           & \multicolumn{4}{c|}{GossipCop}                                                                          \\ \cline{2-9} 
                                                                                & \multicolumn{1}{c|}{F1}       & \multicolumn{1}{c|}{Precision} & \multicolumn{1}{c|}{Recall}   & Accuracy & \multicolumn{1}{c|}{F1}      & \multicolumn{1}{c|}{Precision} & \multicolumn{1}{c|}{Recall}  & Accuracy \\ \hline
GraphConv                                                                       & \multicolumn{1}{c|}{0.78861}  & \multicolumn{1}{c|}{0.755794}  & \multicolumn{1}{c|}{0.79047}  & 0.788201 & \multicolumn{1}{c|}{\textbf{0.90702}} & \multicolumn{1}{c|}{0.96215}   & \multicolumn{1}{c|}{0.94935} & 0.93252  \\ \hline
GATConv                                                                         & \multicolumn{1}{c|}{0.76672}  & \multicolumn{1}{c|}{0.72966}   & \multicolumn{1}{c|}{0.79047}  & 0.766926 & \multicolumn{1}{c|}{0.90483} & \multicolumn{1}{c|}{0.95616}   & \multicolumn{1}{c|}{0.95493} & 0.93186  \\ \hline
GCNConv                                                                         & \multicolumn{1}{c|}{\textbf{0.794199}} & \multicolumn{1}{c|}{0.73147}   & \multicolumn{1}{c|}{0.866664} & 0.794632 & \multicolumn{1}{c|}{0.9022}  & \multicolumn{1}{c|}{0.95372}   & \multicolumn{1}{c|}{0.95536} & 0.93021  \\ \hline
\end{tabular}%
}

\end{table*}
%\FloatBarrier
%================================End of table 6===================

%================Beginning of Table 7=============================
%\FloatBarrier
\begin{table*}[h]
\centering
\caption{Classification scores using Text features}
\label{tab:text-scores}
\resizebox{0.95\textwidth}{!}{%
\begin{tabular}{|cc|cccc|cccc|}
\hline
\multicolumn{1}{|c|}{\multirow{2}{*}{Encoding Technique}}   & \multirow{2}{*}{Truncation type} & \multicolumn{4}{c|}{Politifact}                                                                         & \multicolumn{4}{c|}{GossipCop}                                                                          \\ \cline{3-10} 
\multicolumn{1}{|c|}{}                                      &                                  & \multicolumn{1}{c|}{F1}      & \multicolumn{1}{c|}{Precision} & \multicolumn{1}{c|}{Recall}  & Accuracy & \multicolumn{1}{c|}{F1}      & \multicolumn{1}{c|}{Precision} & \multicolumn{1}{c|}{Recall}  & Accuracy \\ \hline
\multicolumn{1}{|c|}{\multirow{3}{*}{all-mpnet-base-v2}}    & First 512                        & \multicolumn{1}{c|}{0.85044} & \multicolumn{1}{c|}{0.81605}   & \multicolumn{1}{c|}{0.8619}  & 0.85032  & \multicolumn{1}{c|}{\textbf{0.85088}} & \multicolumn{1}{c|}{0.88491}   & \multicolumn{1}{c|}{0.9339}  & 0.85615  \\ \cline{2-10} 
\multicolumn{1}{|c|}{}                                      & Last 512                         & \multicolumn{1}{c|}{0.85044} & \multicolumn{1}{c|}{0.81179}   & \multicolumn{1}{c|}{0.87142} & 0.85027  & \multicolumn{1}{c|}{0.84626} & \multicolumn{1}{c|}{0.88537}   & \multicolumn{1}{c|}{0.92489} & 0.85055  \\ \cline{2-10} 
\multicolumn{1}{|c|}{}                                      & First 256 Last 256               & \multicolumn{1}{c|}{\textbf{0.87043}} & \multicolumn{1}{c|}{0.848}     & \multicolumn{1}{c|}{0.87619} & 0.8715   & \multicolumn{1}{c|}{0.84859} & \multicolumn{1}{c|}{0.87699}   & \multicolumn{1}{c|}{0.94549} & 0.85648  \\ \hline
\multicolumn{1}{|c|}{\multirow{3}{*}{all-distilroberta-v1}} & First 512                        & \multicolumn{1}{c|}{0.8418}  & \multicolumn{1}{c|}{0.80286}   & \multicolumn{1}{c|}{0.8619}  & 0.84167  & \multicolumn{1}{c|}{0.83795} & \multicolumn{1}{c|}{0.86042}   & \multicolumn{1}{c|}{0.96309} & 0.85187  \\ \cline{2-10} 
\multicolumn{1}{|c|}{}                                      & Last 512                         & \multicolumn{1}{c|}{0.84329} & \multicolumn{1}{c|}{0.81243}   & \multicolumn{1}{c|}{0.8619}  & 0.8438   & \multicolumn{1}{c|}{0.83589} & \multicolumn{1}{c|}{0.87144}   & \multicolumn{1}{c|}{0.9339}  & 0.84364  \\ \cline{2-10} 
\multicolumn{1}{|c|}{}                                      & First 256 Last 256               & \multicolumn{1}{c|}{0.86712} & \multicolumn{1}{c|}{0.8375}    & \multicolumn{1}{c|}{0.88571} & 0.86734  & \multicolumn{1}{c|}{0.83935} & \multicolumn{1}{c|}{0.87656}   & \multicolumn{1}{c|}{0.92961} & 0.84891  \\ \hline
\multicolumn{1}{|c|}{\multirow{3}{*}{all-MiniLM-L12-v2}}    & First 512                        & \multicolumn{1}{c|}{0.83946} & \multicolumn{1}{c|}{0.81125}   & \multicolumn{1}{c|}{0.84761} & 0.84761  & \multicolumn{1}{c|}{0.83808} & \multicolumn{1}{c|}{0.87606}   & \multicolumn{1}{c|}{0.92832} & 0.8443   \\ \cline{2-10} 
\multicolumn{1}{|c|}{}                                      & Last 512                         & \multicolumn{1}{c|}{0.82649} & \multicolumn{1}{c|}{0.79331}   & \multicolumn{1}{c|}{0.83809} & 0.82678  & \multicolumn{1}{c|}{0.83517} & \multicolumn{1}{c|}{0.87243}   & \multicolumn{1}{c|}{0.93047} & 0.84233  \\ \cline{2-10} 
\multicolumn{1}{|c|}{}                                      & First 256 Last 256               & \multicolumn{1}{c|}{0.83733} & \multicolumn{1}{c|}{0.81724}   & \multicolumn{1}{c|}{0.83333} & 0.83741  & \multicolumn{1}{c|}{0.84094} & \multicolumn{1}{c|}{0.8824}    & \multicolumn{1}{c|}{0.92103} & 0.84529  \\ \hline
\multicolumn{2}{|c|}{doc2Vec}                                                                  & \multicolumn{1}{c|}{0.74438} & \multicolumn{1}{c|}{0.69244}   & \multicolumn{1}{c|}{0.79047} & 0.74574  & \multicolumn{1}{c|}{0.7239}  & \multicolumn{1}{c|}{0.78986}   & \multicolumn{1}{c|}{0.96952} & 0.778801 \\ \hline
\end{tabular}%
}

\end{table*}
%\FloatBarrier
%================================End of table 7===================

%================Beginning of Table 8=============================
%\FloatBarrier
\begin{table*}[h]
\centering
\caption{Classification scores using fusion techniques}
\label{tab:fusion-scores}
\resizebox{0.95\textwidth}{!}{%
\begin{tabular}{|c|cccc|cccc|}
\hline
\multirow{2}{*}{Fusion Technique} & \multicolumn{4}{c|}{Politifact}                                                                                        & \multicolumn{4}{c|}{GossipCop}                                                                          \\ \cline{2-9} 
                                  & \multicolumn{1}{c|}{F1}       & \multicolumn{1}{c|}{Precision} & \multicolumn{1}{c|}{Recall}   & Accuracy              & \multicolumn{1}{c|}{F1}      & \multicolumn{1}{c|}{Precision} & \multicolumn{1}{c|}{Recall}  & Accuracy \\ \hline
Early Fusion                      & \multicolumn{1}{c|}{0.871431} & \multicolumn{1}{c|}{0.853061}  & \multicolumn{1}{c|}{0.871426} & 0.871642              & \multicolumn{1}{c|}{\textbf{0.93632}} & \multicolumn{1}{c|}{0.95681}   & \multicolumn{1}{c|}{0.96051} & 0.93647  \\ \hline
Late Fusion - Mean                & \multicolumn{1}{c|}{0.885229} & \multicolumn{1}{c|}{0.873164}  & \multicolumn{1}{c|}{0.88095}  & 0.88649               & \multicolumn{1}{c|}{0.93111} & \multicolumn{1}{c|}{0.96386}   & \multicolumn{1}{c|}{0.97296} & 0.95128  \\ \hline
Late Fusion - Classifier          & \multicolumn{1}{l|}{\textbf{0.914531}}         & \multicolumn{1}{l|}{0.891666}          & \multicolumn{1}{l|}{0.940691}         & \multicolumn{1}{l|}{0.917442} & \multicolumn{1}{c|}{0.93011} & \multicolumn{1}{c|}{0.9569}    & \multicolumn{1}{c|}{0.95567} & 0.93245  \\ \hline
\end{tabular}%
}

\end{table*}
%\FloatBarrier
%================================End of table 8===================

For the purpose of classification using graph-level features, we experimented with traditional machine learning algorithms such as ensemble methods, logistic regression and stochastic gradient descent. Specificaly, following algorithms are considered - PassiveAggressiveClassifier, RidgeClassifier, LogisticRegression, SGDClassifier, ExtraTreesClassifier and RandomForestClassifier. The performance of different classifiers using graph-level features on Politifact and GossipCop is illustrated in Table \ref{tab:graph-level-scores}.

We performed comprehensive experiments on the considered dataset to gauge effectiveness of different modalities (Text features and context features) in classification of fake news. Four sets of experimentation are performed as provided below:

\begin{enumerate}
    \item Classification based on manually extracted Graph-level features
    \item Automatic Graph-level classification (GNNs directly applied on propagation graph of news)
    \item Classification based on Content (i.e Text) features of news
    \item Mixed model classification, i.e. combination of both text based and propagation based features
\end{enumerate}

Classifiers could not perform that well on Politifact with RandomForestClassifier reaching maximum f1-score of 0.48. The likely reason for this is the small size of Politifact dataset which does not allow for learning of meaningful patterns. However, a decent f1-score of 0.88 is reached on GossipCop dataset. In both cases, RandomForestClassifier performs best closely followed by ExtraTreesClassifier.

Using Graph Neural Networks, we achieve a maximum f1-score of 0.79 on Politifact with GCNConv and 0.907 on GossipCop with GraphConv. We experimented with different number of layers and embeddings size for convolutional and found that 4 layers and 64 dimensions provides comparable results to other settings while having a shorter training time. Results are shown in Table \ref{tab:gnn-results}.

Text content of a news article provide important signals in distinguishing fake and real news. The results on this modality using different models as discussed in section methodology is provided in Table \ref{tab:text-scores}. When classifying using text content, we consider a specific part of text input if it is longer than maximum sequence length of 512 tokens. In case of Politifact dataset, we found that using combination of first 256 and last 256 provides significantly better results than considering first 512 or last 512 tokens as shown in Table \ref{tab:text-scores}. Similar but less pronounced trend can be observed in GossipCop as well. This can be attributed to fact that taking first 256 and last 256 tokens effectively summarizes the entire news article. \textit{all-mpnet-base-v2} outperforms \textit{all-distilroberta-v1} and \textit{all-MiniLM-L12-v2} on all truncation techniques on both datasets. At cost of slight decrease in accuracy, \textit{all-distilroberta-v1} computes the embeddings in half the time as  \textit{all-mpnet-base-v2}.

Proposed fusion techniques perform significantly better than unimodal models (text and GNN), as shown in Table \ref{tab:fusion-scores}. All fusion techniques achieved significant improvements over baseline text and GNN models when using GossipCop dataset. An improvement of 3\% is achieved over best performing unimodal model. On Politifact, Late Fusion with mean aggregation performs sligthly better than Early fusion but has the advantage of having significantly lesser training time than classifier approach while Late Fusion with classifier outperforms both fusion and baseline model by atleast 3\%. Likely conclusion that can be derived from these reuslts is that Late Fusion Mean can be used as a good trade-off between accuracy and computational expense when training data is sparse. Since there was no significant difference between f1-scores of fusion techniques on GossipCop, Early fusion is the best choice here because of low training time compared to other fusion techniques.

\section{Conclusion and Future Scope}

In this work, we explored a method of detecting fake news based on its propagation characteristics on social media and its content. Experiments were performed to demonstrate that using both content and propagation characteristics provides better performance than relying on a single modality. Fake news detection will be most beneficial when fake news can be identified at an early propagation stage. GNN approaches that can learn on structured data like graphs seem to be promising for investigating such directions. Moreover, most past approaches used in fake news detection are not interpretable. Hence, exploring the proposed model for interpretation and explanation of the achieved results can be another future direction of our research.

% % % use section* for acknowledgment
% \section*{Acknowledgment}

% The authors would like to thank... (if any external persons/agency helped you for performing experimentation and gathering the data)

\bibliographystyle{IEEEtran}
\bibliography{reference}

\end{document}